# Two-stage Hydrogen Compression Using Zr-based Metal Hydrides

Evangelos D. Koultoukis[1,2,a], Sofoklis S. Makridis[1,b],
Daniel Fruchart[3,c] and Athanasios K. Stubos[1,d]

[1]Institute of Nuclear Technology and Radiation Protection, NCSR "Demokritos",
Ag.Paraskevi, 15-310, Athens, Greece

[2]McPhy Energy S.A., Z.A. Quartier Riétière, 26190 La Motte-Fanjas, France

[3]Néel, CNRS, 25 Avenue des Martyrs, BP 166, 38042 Grenoble Cedex 9, France

[a]v.koultoukis@ipta.demokritos.gr, [b]sofmak@ipta.demokritos.gr, [c]daniel.fruchart@grenoble.cnrs.fr, [d]stubos@ipta.demokritos.gr

**Abstract.** Zr-based $AB_2$-Laves phase type alloys containing the same type of A and B metals, have been prepared from pure elements by melting and subsequent re-melting under argon atmosphere by using a HF-induction levitation furnace. Characterization of the alloys has resulted from powder X-Ray Diffraction (XRD) measurements and SEM/EDX analyses. Systematic PCI (Pressure-Composition-Isotherms) measurements have been recorded at 20 and 90 °C, using a high-pressure Sievert's type apparatus. The purpose of this study is to find a series of alloys promptly forming metal hydrides (MH) with suitable properties in order to build a MH-based hydrogen compressor, working in the same way between 20 and ~100 °C.

**Introduction**

The development of lightweight high-pressure hydrogen storage vessels has led the working pressure to be much higher than it used to be. As a result, a need for efficient, safe, and low cost hydrogen compressors has begun to emerge.

Non-mechanical hydrogen compressors have several advantages over the mechanical ones, including smaller size, less expensive, lower operating and maintenance costs. Since the hydrogen absorption-desorption plateau pressure of a metal hydride (MH) varies with temperature according to the van't Hoff equation,

$$\ln P = \Delta H/RT - \Delta S/R, \qquad (1)$$

the MH compressors are thermally powered systems that use the ability of reversible metal hydrides to compress hydrogen without any contamination [1]. They also provide the ability to connect them to the outlet of electrolysers [2]. Moreover, using heat wastes to feed the chemical compressor will enhance the overall efficiency of the system [3].

**Experimental**

All the alloys were prepared under pure argon atmosphere (5N5) by using the HF-induction levitation melting method in a cold crucible. During the melting process, the ingots were turned over and re-melted 2 times at least in order to ensure homogeneity. X-ray Diffraction was performed to investigate the phase structure of experimental alloys. The diffraction data were collected at room temperature on a Philips CubiX-XRD device in the Bragg-Brentano geometry at

$\lambda(Cu_{K\alpha}) = 0.15418$ nm using the following conditions of $10 < \theta_B < 100$ degrees and a counting step of 0.02 degrees. In order to deliver optimized information on the crystal structure modifications due to hydrogenation, Rietveld analysis (RIETICA) has been performed on the as-cast powders as well as on the hydrogenated powders. The microstructure of each sample was investigated by using a FESEM Zeiss Ultra Plus electronic microscope. The hydrogenation dynamic and thermodynamic properties were evaluated using a volumetric high pressure Sievert's-type device. All PCI traces were recorded at 20 and 90 °C, respectively. All samples have been loaded into the sample holder at room temperature and under ambient atmosphere. To ensure the best degassing procedure, the samples were kept under vacuum at 350 °C for at least 30 min. This type of alloys though, requires a high energy activation procedure in order to react effectively with hydrogen. For this reason, 8 activation cycles have been performed using the following procedure: Once the temperature of the sample has reached 350 °C following the previous out-gassing step, a pressure of 40 bars of hydrogen gas was admitted into the sample holder. Then temperature was decreased approximately up to 2-5 °C and the sample was left to absorb hydrogen until equilibrium was reached. After that, the temperature was increased once more up to 350 °C and we waited until pressure equilibrium was reached again. This procedure repeated for 8 times, while the hydrogen pressure was always applied into the sample holder.

**Results**

Some of the Rietveld-analysed XRD patterns are shown on Fig. 1. All compounds appear to be single phase, corresponding to either C14 or C15 of the $AB_2$ Laves series as reported for the $Zr_{0.5}Ti_{0.5}Fe_{1.2}Ni_{0.4}V_{0.4}$ and $Zr_{0.75}Ti_{0.25}Fe_{1.0}Ni_{0.8}V_{0.2}$, respectively. The single-phase samples are always analysed in the $AB_2$ Zr-based intermetallic compounds during improvement of the kinetic of hydrogenation properties [4,5]. It has been observed that no significant change has occured on the Full Width at Half Maximum Height of the profile of the XRD peaks after the PCI's measurements. This means that the alloys are resistant to disproportionation after applying several cycles under high hydrogen pressure (>100 bars). Resistant formula are highly desirable since metal hydride alloys are expected to work for several $10^3$ absorption/desorption cycles, so long life reversible cycles are requested. All crystal structure data extracted from the Rietveld analysis are displayed on Table 1 and 2.

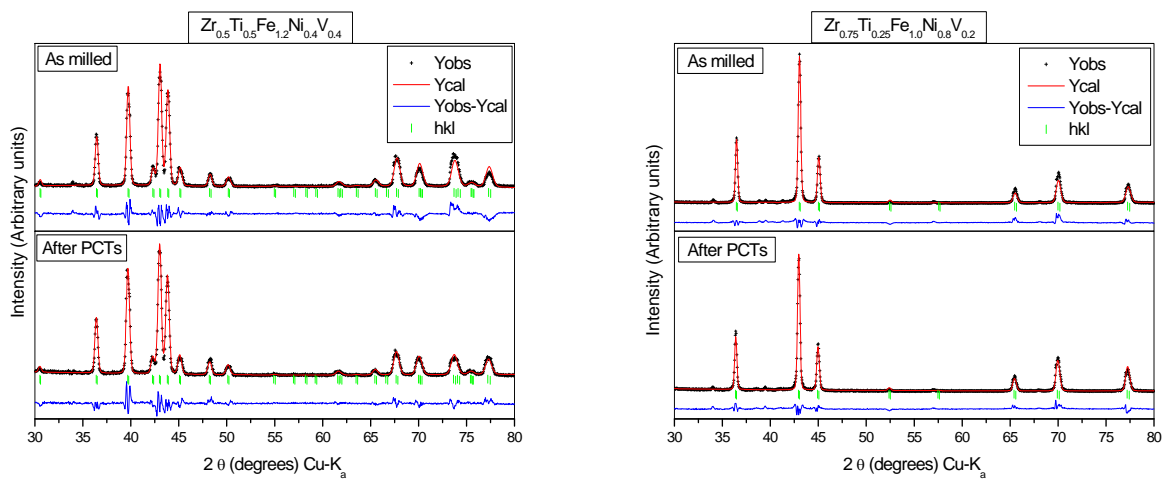

**Fig. 1. Rietveld analysis of $Zr_{0.5}Ti_{0.5}Fe_{1.2}Ni_{0.4}V_{0.4}$ and $Zr_{0.75}Ti_{0.25}Fe_{1.0}Ni_{0.8}V_{0.2}$, respectively.**

**Table 1. Results of a Rietveld-type analysis of $Zr_{0.5}Ti_{0.5}Fe_{1.2}Ni_{0.4}V_{0.4}$**

| Sample $Zr_{0.5}Ti_{0.5}Fe_{1.2}Ni_{0.4}V_{0.4}$ | Space Group (No.) | Phase | Unit cell parameters (Å) a | c | Unit cell volume (Å$^3$) | Phase abundance (wt%) | $R_{Bragg}$ |
|---|---|---|---|---|---|---|---|
| As milled | P 6$_3$/mmc | C14 | 4.9362 | 8.0421 | 169.7088 | - | 7.82 |
| After PCTs | P 6$_3$/mmc | C14 | 4.9390 | 8.0463 | 169.9854 | - | 5.09 |

| | Refinement parameters | | |
|---|---|---|---|
| | $R_p$ | $R_{wp}$ | $\chi^2$ |
| As milled | 15.4 | 11.4 | 0.60 |
| After PCTs | 12.3 | 11.5 | 0.50 |

**Table 2. Rietveld analysis results of the $Zr_{0.75}Ti_{0.25}Fe_{1.0}Ni_{0.8}V_{0.2}$ alloy**

| Sample $Zr_{0.75}Ti_{0.25}Fe_{1.0}Ni_{0.8}V_{0.2}$ | Space Group (No.) | Phase | Unit cell parameters (Å) a | c | Unit cell volume (Å$^3$) | Phase abundance (wt%) | $R_{Bragg}$ |
|---|---|---|---|---|---|---|---|
| As milled | F d -3 m | C15 | 6.9922 | | 341.861 | - | 5.02 |
| After PCTs | F d -3 m | C15 | 6.9888 | | 341.356 | | 4.98 |

| | Refinement parameters | | |
|---|---|---|---|
| | $R_p$ | $R_{wp}$ | $\chi^2$ |
| As milled | 15.1 | 6.9 | 0.73 |
| After PCTs | 13.3 | 7.7 | 0.58 |

The SEM/EDX micrographs have been recorded on polished samples using up to 3 µm corindon pasta. Fig. 2 presents the microstructure of the two selected alloys. It is worth to note that the $Zr_{0.5}Ti_{0.5}Fe_{1.2}Ni_{0.4}V_{0.4}$ sample crystallize homogeneously with a typical fir-tree type structure, as most AB$_2$ alloys having a C14 Laves phase type, while such a typical structure is absent for the second alloy, probably to relate to the higher amount of both Zr and Ni. Anyway, the two alloys exhibit a very good homogeneity which is a result of the rapid solidification melting technique after quenching in the water cooled cold crucible.

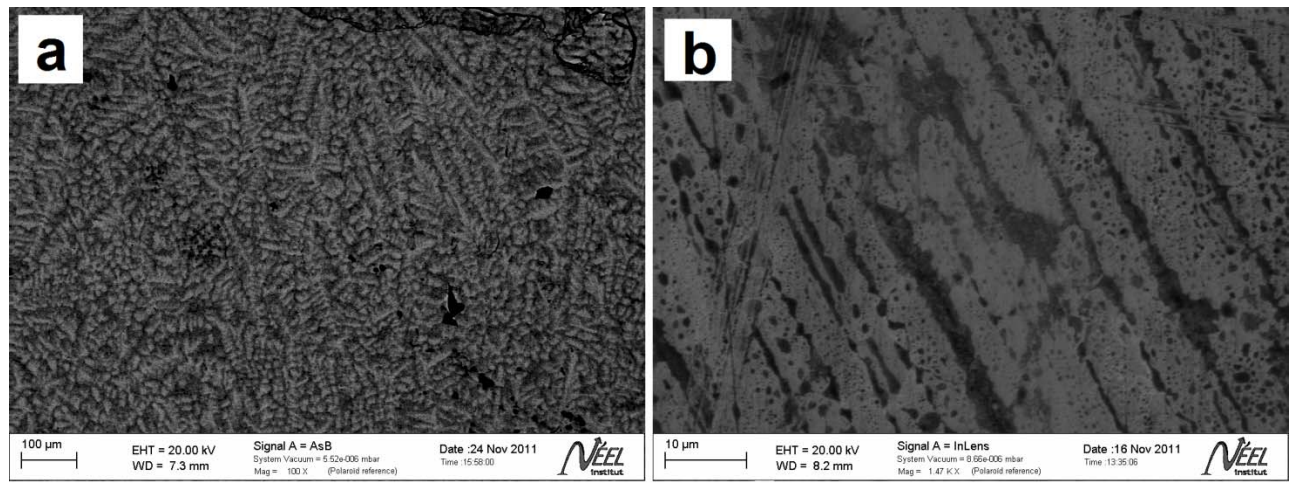

**Fig. 2.** SEM micrographs of $Zr_{0.5}Ti_{0.5}Fe_{1.2}Ni_{0.4}V_{0.4}$ and $Zr_{0.75}Ti_{0.25}Fe_{1.0}Ni_{0.8}V_{0.2}$, respectively.

PCI traces have been recorded at 20 and 90 °C, under pressure up to 100 bars. They are shown on Fig. 3, left and right respectively. As can be seen, hydrogen pressure can be increased up to 10 times in a two stage assembly, by using such different alloy compositions. For the first sample, which is the lowest pressure alloy, the input conditions should be 10 bars hydrogen pressure at 20 °C at absorption, while the corresponding hydride can release almost 0.8 wt.% of hydrogen under 40 bars when temperature reaches 90 °C. This final pressure would be the input pressure at room temperature for the second sample, the highest pressure alloy. Due to software and hardware limitations of the used PCI apparatus, the maximum hydrogen pressure that could be used for the isotherm trace measurements was established to 100 bars. And for such a reason, the desorption curve of the second compound was not completed. Since the second hydrogenated alloy seems to desorb mostly hydrogen under more than 100 bars at 90 °C, several isotherms have been also performed at lower temperatures, so that the equilibrium pressure ranging in the 0-100 bars interval. Thus, using these sets of experimental data, application of the van't Hoff equation has let us to fix the equilibrium pressure of the second compound at more than 90 °C. In such a way, the $Zr_{0.75}Ti_{0.25}Fe_{1.0}Ni_{0.8}V_{0.2}$ hydride would release hydrogen at about 105 bars, as can be anticipated from Fig. 3. Cycling tests of the hydrogenated alloys operated under constant pressure and temperature has showed that the kinetics are very fast. Maximum hydrogen uptake is reached in less than ~ 20 min (that is more than 1.6 wt.% and 1.2 wt.%, respectively) with almost 80% of the maximum uptake taking place in the first 5 min hydrogen pressure exposure. The hydrogen desorption process appears even faster at the same applied temperatures, reaching equilibrium in almost 8 min. Indeed, both these behaviors depend as well on the used system, i.e. the efficiency of the thermal transfers.

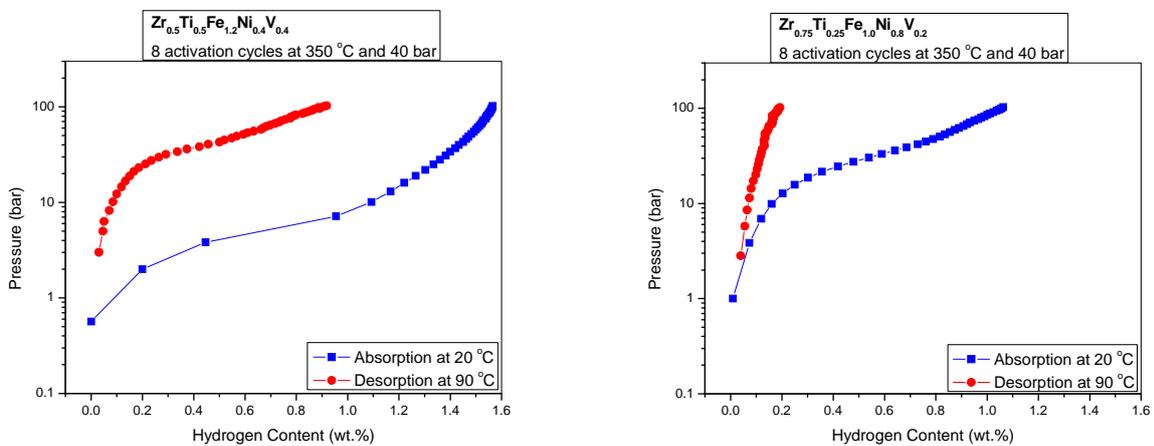

**Fig. 3.** PCI of $Zr_{0.5}Ti_{0.5}Fe_{1.2}Ni_{0.4}V_{0.4}$ and $Zr_{0.75}Ti_{0.25}Fe_{1.0}Ni_{0.8}V_{0.2}$, respectively.

**Discussion and Conclusion**

In summary, the structure and microstructure, and the thermodynamics and kinetics of hydrogenation of two $AB_2$ Zr-based compounds have been investigated. It was found that the crystal structures correspond to the C14- and C15-types of Laves phases for $Zr_{0.5}Ti_{0.5}Fe_{1.2}Ni_{0.4}V_{0.4}$ and $Zr_{0.75}Ti_{0.25}Fe_{1.0}Ni_{0.8}V_{0.2}$, respectively, which could play a role in determining higher absorption/desorption plateau pressures for the second compound. In fact, the volume of the C15 unit cell measured here is only slightly larger than that of the ideal C36 $AB_2$-polytype structure. Ideally it must be twice, and the very small difference as found here is explained in term of the Zr/Ti proportion varying from 1 to 3 in the A sites for C14 and C15, so the mean volume of the

Friauf's polyhedron being very little larger in the cubic than in the hexagonal system, respectively. In fact, the main difference in the equilibrium pressure of plateaus as seen experimentally here does result of a volume effect but more effectively on the chemical attraction to hydrogen of the metal elements. Since Zr and Ti have very similar relative electronegativities (1.4 to 1.5 in the Pauling's scale), one can anticipate that the nature and the proportion of metal elements Fe, Ni and V, plays the major role in establishing the relative stability of their respective hydrides via both the electronegativity differences and more probably the filling their 3d bands.

On more extrinsic aspects, it is shown that both compounds show a very good disproportionation resistance after several activation and hydrogenation cycles, which is a desired property. Cycling tests have proven that both alloys react very fast with hydrogen, after a moderate activation procedure has been realized. Similarly, the hydrides enable release all absorbed hydrogen, with almost no hysteresis effect, also the late being a very important aspect for the planed application. Both these favorable peculiarities (no disproportionation, almost no hysteresis) could be related to the nice homogeneity of the samples in terms of as-received microstructure.

Finally, using two parent formuled-polytype of $AB_2$ compounds, the hydrogen compression ratio reaches almost 10, delivering an output pressure of more than 100 bars. The expected hydrogen transfer should be close to 1 wt.% when processing between 20 to 90 °C using a ~180 g batch of alloy per step, thus corresponding to little less than 20 liters $H_2$ transferred per step in about 20 min in the present but not yet optimized heat transfer conditions. A 3 kg alloys containing step (volume ~ 0.5 liter) should permit compression, at *minimum minimorum*, 1 $Nm^3$ $H_2$ gas per hour (more than 20 kg $H_2$ a day).

In forthcoming papers, the crystal structure of the compounds and their hydrides (deuterides) will be discussed for and compared to those of other parents compounds planed to complete the ability working in the highest compression side (200 up to 300 bars). Also a band structure calculation approach will be detailed in order to assert the here above first conclusions, and to demonstrate the powerful efficiency of *ab initio* considerations in formulating the $(Zr,Ti)TM_2$ compounds with determined stability of their hydrides on demand.

## Acknowledgements

This work is supported by the ATLAS-H2 European Project PIAP-GA-2009-251562 that grants our R&D activities.